\documentstyle[12pt]{article}

\begin{document}
 
\begin{titlepage}

\begin{center}
 
\vspace{.5 true  cm}
 
\Large
{\bf Renormalization Group Approach to the}\\
\vspace{6pt}
{\bf \mbox{Normal State of Copper-Oxide Superconductors} }\\
\vspace{6pt}

\vspace{1.5 true  cm}
 
\large
{\bf J. Gonz\'alez\S\footnote{e-mail: emgonzalez@iem.csic.es},
F. Guinea\dag \footnote{e-mail: paco@ccuam3.sdi.uam.es} and
M.A.H. Vozmediano\ddag }
 
\vspace{0.5 true  cm}
 
\normalsize
\S{\em Instituto de Estructura de la Materia, CSIC} \\
{\em Serrano 123, 28006 Madrid, Spain} \\
 
\dag{\em Instituto de Ciencia de Materiales, CSIC} \\
{\em Cantoblanco, 28049 Madrid, Spain.} \\
 
\ddag{\em Departamento de Matem\'aticas, Universidad Carlos
III de Madrid} \\
{\em Avda. Butarque 15, 28913 Legan\'es (Madrid), Spain.}
 
\vspace{.5 true  cm}

\end{center}
 
\begin{abstract}

We study by means of renormalization group techniques the 
effect that on the two-dimensional electron liquid may have
the van Hove singularities observed experimentally in the
copper-oxide superconductors. We find significant deviations 
from Fermi liquid behavior, that lead to the appearance of 
an unstable fixed point in the renormalization group flow of
the effective coupling constant. Besides the attenuation of 
electron quasiparticles already known on phenomenological 
grounds, our approach is able to explain the reduction in
the dispersion of the band as well as the pinning of the
Fermi level near the singularity, as observed in the 
photoemission experiments.

\vspace{3 true cm}
 
\end{abstract}

\vspace{3 true cm}
\vskip5cm
\noindent
 
\newpage
 
\end{titlepage}
\section{Introduction}

During recent years there has been great interest in finding
theoretical models displaying the so-called ``non-Fermi'' liquid
behavior\cite{nf}. 
This has not been a trivial search, mainly due to the
fact that the Fermi liquid framework is quite robust in the
context of interacting electron systems. The Fermi liquid
concept based on Landau's theory acts as the most valuable
paradigm in the description of real three-dimensional
metals\cite{landau}. 
The relevance of such paradigm is not constrained to situations
in which the interaction strength is weak and goes beyond the
range of applicability of perturbation theory. For some reasons,
not quite well understood until recently, there are not many
ways in which an electron system, in spatial dimension $D > 1$, 
may be destabilized out of the Fermi liquid picture.
Some of the most outstanding real examples in which that happens
are furnished by the copper-oxide compounds that become
superconductors at high transition temperatures. In the metallic
regime of these materials, some of the most prominent
experimental properties (low-temperature resistivity,
specific heat) have an abnormal behavior that cannot be
explained by Fermi liquid theory\cite{varma}. 
Thus, for the copper-oxide
compounds it is not only a question of explaining the unusual
superconducting transition temperatures, but also of
understanding the unconventional properties of the normal state
---what may serve furthermore to uncover the superconducting
mechanism. 

The most modern approach to a general theory of interacting
fermion systems has put forward the idea that the Fermi liquid
describes a universality class in two and three spatial
dimensions\cite{sh,pol}, 
much in the same spirit that the Luttinger liquid is
the class for a wide set of one-dimensional electron systems\cite{hal}. 
In order to apply the renormalization group methods to many-body
fermion systems, the main achievement has been a careful
analysis of the interactions on the grounds of the usual
classification of the perturbations of a statistical 
theory\cite{sh,pol}
---that is, relevant, irrelevant and marginal operators. 
Unlike for the one-dimensional fermion systems, this
is not straightforward in higher dimensions, since the
interaction involves in general excitations attached to
different points of the Fermi surface. Thus, one has to talk
properly of the existence of a manifold of coupling constants,
not just of a finite number of them. Under very general
assumptions, like the regularity of the interaction potential in
that manifold of couplings, it turns out that there are
essentially two different marginally relevant perturbations of
the fermion system, corresponding to the superconductivity
instability, which is always present, and to the charge density
or spin density instability, which arises when the Fermi surface
has the so-called nesting property. The rest of marginal
interactions translates into the familiar $F$ function of
Landau's Fermi liquid theory\cite{sh}. Then, in this setup there is
little room left to the description of additional deviations
from Fermi liquid behavior, which may be conceived only at the
expense of relaxing some of the initial assumptions. 

Confronting such theoretical framework, there is clear
experimental evidence that most of the copper-oxide
superconductors have normal state properties incompatible with
those of a Fermi liquid. The most notorious of them is the
linear resistivity as a function of the temperature (at the
optimal doping for the superconductivity), while a quadratic
dependence is the only behavior that can be accounted for in
Fermi liquid theory\cite{book}. 
On the other hand, the origin of the
superconductivity in the cuprate oxides remains unexplained,
although there is strong experimental evidence in favor of a
mechanism based on electron-electron interactions. The electrons
responsible for the conductivity in the normal state, and for
the appearance of the superconducting condensate, are supposed
to be localized on weakly coupled planes. The existence of
important correlation effects has been used to propose, in
the context of prototypes like the Hubbard or $t$-$J$ models, 
a novel two-dimensional electron liquid state, that
should be responsible both for the unusual normal state
properties as well as for the high-$T_c$ 
superconductivity\cite{novel}. In
the doped materials, however, there is nowadays clear evidence
of the existence of a Fermi surface typical of a metallic 
band\cite{band,photo}.
Furthermore, the experiments based on angle-resolved
photoemission spectroscopy have shown interesting features of
the dispersion relation, like almost dispersionless bands and
extended van Hove singularities (saddle points) near the Fermi
level\cite{photo,abr}. 
Based on this experimental evidence, supplemented 
by phenomenological arguments, it has been proposed that the 
unusual properties of electrons near such  singular points are 
the cause of the superconductivity in these systems\cite{Newns}.
Although a variety of theoretical approaches have been 
attempted\cite{hove,mark,tsuei,patt}, 
a full treatment of interacting electrons 
near a saddle point in a 2D dispersion relation remains to be done.

The aim of this paper is to study the behavior of the
two-dimensional electron liquid near a van Hove singularity by
means of a renormalization group approach. Due to the presence
of the saddle point in the dispersion relation, the usual
description of the fermion system breaks down  from the start.
In the standard many-body treatment as well as in the modern
renormalization group approach, the fact that one can linearize
the dispersion relation at each point of the Fermi surface plays
a crucial role. When that is possible, one may disregard the
particular features of the Fermi surface. Thus, in the usual
many-body approach one does not care about its fate under
quantum corrections, since all its points are regarded as
essentially equivalent. When there is a van Hove singularity at
or very close to the Fermi surface that is no longer true.
Physically this can be easily understood since at the point of
the singularity the density of states diverges logarithmically.
From a technical point of view, the gradient of the dispersion
relation at the saddle point vanishes, what makes unclear that
the traditional many-body approach can be followed. Quite
remarkably, however, the logarithmic divergences associated to a
two-dimensional van Hove singularity make the problem specially
suitable to renormalization group schemes, along the lines
discussed in \cite{sh,pol}.

In the following, we implement such a RG treatment of interacting 
electrons near a 2D van Hove singularity. The model to be studied 
is presented in the next section. Then, the renormalization of the 
coupling constant is worked out in section 3.
In section 4 we discuss how the renormalized potential influences 
the quasiparticle properties. The renormalization group flow of
the model is presented in 
section 5. A discussion of the physical consequences is given in
section 6. The main conclusions are drawn in section 7.

\section{Many-body theory of the 2D van Hove singularity}

A van Hove singularity is a saddle point in the dispersion relation
of the electron states $\varepsilon ( {\bf k} )$. In its vecinity, the
density of states diverges logarithmically, in two dimensions, 
and shows cusps in three dimensions. 
The logarithmic divergence leads to a singular screening of the
interactions, in the same way as for the 1D Luttinger liquid\cite{Lutt}.

The copper-oxygen planes where the electrons near the Fermi level
reside have an almost perfect tetragonal symmetry. Thus, the
crystal wavevector ${\bf k}$ is defined in a square Brillouin Zone,
such that $- \pi < k_x , k_y < \pi$. Experiments show that the
Fermi surface lies close to the high symmetry points $( 0 , \pm \pi )$
and $( \pm \pi , 0 )$. Thus, we expect two inequivalent van Hove
^^ ^^ points" at the edges of the Brillouin Zone, as depicted in
Fig. 1 (note that
the points previously defined, are, by symmetry, extrema of the
dispersion relation).

We can obtain a reasonable description of the low-energy 
properties of the system by expanding the dispersion relation around
these points.  Denoting them by $A$ and $B$, and shifting the
origin of momenta to each of these points, we can write, in general,

\begin{equation}
\varepsilon_{A,B} ( {\bf k} ) \approx ( t \pm \Delta t ) k_x^2
- ( t \mp \Delta t ) k_y^2
\end{equation}
where $\Delta t$ is proportional to the deviation of the
angle between the two separatrices from $\pi / 2$.

Thus, the problem can be mapped onto a model with electrons
with two flavors, which denote each of the two inequivalent singularities.
The full complexity of the interacting system is already present
if we neglect this flavor degeneracy (which can be achieved, for instance, 
by means of an orthorhombic distortion which splits the two points).
In the following, we will consider a system with electrons near
a single singularity. Intersingularity (Umklapp) effects will
be addressed in future work.

For a single singularity, a suitable scaling of lengths  leads
to the simpler dispersion relation

\begin{equation}
\varepsilon ( {\bf k} ) = t(k_x^2 - k_y^2)
\label{disp}
\end{equation}
We describe the situation in which the Fermi energy is
right at the level of the van Hove singularity. The case in
which the Fermi level deviates slightly from the position of the
singularity can be treated by means of a small shift in the
chemical potential of the system, which just introduces a small
infrared cutoff in the renormalization group approach. There
remains the question, though, about the physical significance of
a situation in which the filling of the Fermi sea stops at the
level of the singularity. According to our point of view, the
shape of the Fermi line may suffer changes due to the
interaction in the system and it is not clear, in principle,
that the Fermi level may find a stable location near the
singularity. We will see later on that this tunning of the Fermi
level to the singularity is actually quite natural, from the point of
view of the renormalization group approach.
 
We take then the chemical potential at $\mu = 0$ for the
dispersion relation (\ref{disp}), so that the levels with
$\varepsilon ({\bf k}) < 0$ are filled and the levels with 
$\varepsilon ({\bf k}) > 0$ are all empty.
The Fermi line is given by the two straight asymptotes $k_x =
\pm k_y$. Of course, ir order to accomplish a proper description
of the statistical system some bandwith cutoff has to be
introduced that regularizes the formally infinite Fermi sea of
(\ref{disp}). We implement in what follows a cutoff in momentum
space $(k_x , k_y)$ that preserves the SO(1,1) symmetry of the
dispersion relation (\ref{disp}), by picking the states that lie
in the region with $-\Lambda < \varepsilon ({\bf k}) < \Lambda$.
Thus, $\Lambda $ has the character of an energy (or frequency)
cutoff. In fact, this possibility of controlling the energy
scale of the processes in an invariant way is what allows us to
carry out the renormalization program that we apply below. The
behavior of the interactions as we reduce the cutoff $\Lambda $
gives us information about the form of the low-energy effective
theory for the system. In this respect, the procedure is similar
to that adopted in the usual description of the Luttinger
liquid\cite{Lutt} or in the renormalization group approach 
to Fermi liquid theory\cite{sh,pol}. 

The first step in this approach to the low-energy effective
theory of the van Hove singularity is to write down a classical
interacting theory that is scale invariant. We take as starting
point the system of electrons with the dispersion relation
(\ref{disp}) and the typical four-fermion interaction of
density-times-density type. The action of the model reads, for
the simplest case of a local interaction,
\begin{eqnarray}
S  & = &  \int dt' \: d^2 r \sum_{\sigma} \left( 
 i \Psi^{+}_{\sigma} ({\bf r}, t')
\partial_{t'} \Psi_{\sigma }({\bf r}, t') + 
 t \; \Psi^{+}_{\sigma }({\bf r}, t') (\partial^2_x - \partial^2_y )
 \Psi_{\sigma }({\bf r}, t') \right)       \nonumber   \\
 &   &   - \frac{U}{2} \int dt' \: d^2 r \;  \sum_{\sigma }
 \Psi^{+}_{\sigma }({\bf r}, t') \Psi_{\sigma }({\bf r}, t')
  \;  \sum_{\sigma' }
 \Psi^{+}_{\sigma' }({\bf r}, t') \Psi_{\sigma' }({\bf r}, t') 
\label{act}
\end{eqnarray}
$\Psi_{\sigma }({\bf r}, t)$ is the electron field made out of
all the modes that lie in the region determined by the cutoff,
$\sigma $ being the spin index $\sigma = \uparrow, \downarrow$,
\begin{equation}
\Psi_{\sigma }({\bf r}, t) = \frac{1}{(2 \pi)^3} \int_{\cal S}
d\omega \: d^2 k \; e^{-i \omega t + i {\bf k}\cdot {\bf r}} 
 \; a_{\sigma}
({\bf k}, \omega )  \;\;\;\;\;\; {\cal S} = \{ {\bf k} \mid 
-\Lambda < \varepsilon ({\bf k}) < \Lambda \}
\end{equation}

The scaling of the different terms in (\ref{act}) is best
analyzed by passing to the frequency-momentum $(\omega , {\bf k})$
representation, in which the action reads
\begin{eqnarray}
S  & = & \int d \omega d^2 k \sum_{\sigma} \left( \omega 
\; a^{+}_{\sigma}({\bf k}, \omega ) a_{\sigma}({\bf k}, \omega )
- t (k^2_x - k^2_y) \; a^{+}_{\sigma}({\bf k}, \omega ) 
   a_{\sigma}({\bf k}, \omega ) \right)   \nonumber   \\
  &   &  - \frac{U}{2} \int d\omega d^2 k \; 
\left( \rho_{\uparrow } ({\bf k}, \omega)  +
      \rho_{\downarrow } ({\bf k}, \omega) \right)
 \: V( {\bf k}) \: 
\left( \rho_{\uparrow } (-{\bf k}, -\omega) +
      \rho_{\downarrow } (-{\bf k}, -\omega) \right)
\label{actk}
\end{eqnarray}
where $\rho_{\sigma }({\bf k}, \omega)$ are the Fourier
components of the density operator $\Psi^{+}_{\sigma} ({\bf r}, t)
\Psi_{\sigma} ({\bf r}, t)$
\begin{equation}
\rho_{\sigma} ({\bf k}, \omega) = 
 \frac{1}{(2 \pi)^3} \int d \omega_p d^2 p
\; a^{+}_{\sigma} ({\bf p} - {\bf k}, \omega_p - \omega )
   a_{\sigma} ( {\bf p},  \omega_p )
\end{equation}
The case of a purely local interaction corresponds to $V( {\bf
k}) = 1$. In what follows we will consider however the more
realistic case in which the interaction has some finite range in
real space. We will then suppose that the potential $V( {\bf
k})$ is constant but has a finite support in ${\bf k}$ space, so
that the discussion can be carried out for a spin-independent
interaction. 

We determine the scaling dimension of the operator $a_{\sigma} 
({\bf k}, \omega )$ in the usual way, by demanding the scale
invariance of the free term in the action. Given the nonlinear
character of the dispersion relation, it turns out that under a
change of scale of the momenta
\begin{equation}
{\bf k} \rightarrow s {\bf k}
\label{sck}
\end{equation}
the frequency has to scale in the form
\begin{equation}
\omega \rightarrow s^2 \omega
\label{sce}
\end{equation}
This different scaling of the frequency and the momentum is
something natural in the model, since to keep up with the
approach to the Fermi level $\varepsilon ({\bf k}) = 0$ the
energy has to be reduced as the square of the reduction in the
momentum scale. From (\ref{sck}) and (\ref{sce}) we determine
the scaling of the electron operator $a_{\sigma}({\bf k},
\omega)$ that renders invariant the first piece of (\ref{actk})
\begin{equation}
a_{\sigma}({\bf k}, \omega) \rightarrow s^{-3} a_{\sigma}({\bf k},
\omega)
\label{sca}
\end{equation}

A most important point is now to check whether by application of
the scaling transformation (\ref{sck}), (\ref{sce}) and
(\ref{sca}) the interaction strength $U$ in the last term of
(\ref{actk}) goes to zero, diverges or remains invariant in the
infrared limit $s \rightarrow 0$. The first case would
correspond to an irrelevant interaction, that should be
neglected in the effective field theory at low energies, while
in the second case the last term in (\ref{actk}) would play the
role of a relevant perturbation of the noninteracting theory,
spoiling the renormalization group approach. We see, however, 
that the actual case is that the interaction remains marginal at
the classical level, i.e. the coupling strength $U$ does not
scale as $s \rightarrow 0$. Under these conditions, the effect
of the quantum corrections in the model turn out to be crucial
since, though small they may be, they are able to destabilize
the naive scaling in one or the other direction. 

The whole issue of the renormalization of the $U$ parameter
becomes of the utmost interest since, as follows from Shankar's
renormalization group approach to fermion systems, the
interaction in Fermi liquid theory (constrained to the forward
scattering channel) is not renormalized\cite{sh}. It has to be pointed
out, however, that the case of a dispersion relation like
(\ref{disp}) becomes special, in the sense that the interaction
is marginal at the classical level irrespective of the
kinematics of the scattering process, as we have seen above.
This enables us to pose the problem in terms of the
renormalization of just one coupling constant $U$, which affects
evenly to all the scattering channels ---apart from the BCS
channel, which does require a particular kinematics.

In what follows we adopt a statistical field theory description
of the model, sticking to a path integral formulation based on
the action (\ref{actk}). The fundamental object in this approach
is the partition function
\begin{equation}
{\cal Z} = \int {\cal D}\Psi \: {\cal D}\Psi^{+} e^{iS(\Psi,
\Psi^{+})} 
\label{pi}
\end{equation}
where, as stated above, we restrict the functional integration
to the fermion modes within energies $-\Lambda$ and $\Lambda$
measured with respect to the level of the singularity. The Fermi
sea that we are considering corresponds to a noncompact region
in momentum space, though the number of states in the region
becomes infinite only in the infinite volume limit. We will see,
for instance, that the four-fermion interaction and the
self-energy correction of the model can be regularized by the
use of just the bandwith cutoff $\Lambda $. The dependence of
the quantum corrections on $\Lambda $ gives us the information
about how the different parameters of the model are renormalized
when the cutoff is reduced from $\Lambda $ to $\Lambda -
d\Lambda $, what in practice is equivalent to perform the
partial integration of a slice of high-energy modes in the path
integral (\ref{pi}). The recursive implementation of this
renormalization group transformation, studied by the flow
as $\Lambda \rightarrow 0$ of the scale dependent parameters,
leads to the effective field theory governing the
low-energy physics of the model.

\section{Coupling constant renormalization}

We study first the renormalization of the interaction potential
$V( {\bf k})$, which gets the quantum corrections shown by the
diagram in Fig. 2, to second order in perturbation theory.
The figure represents the contribution to the corresponding
vertex function $i \Gamma ({\bf k}, \omega)$, i.e. the part of
the interaction without the external legs.
\footnote{We do not mix up at this point the Cooper channel, 
which requires a particular kinematics, with the generic
corrections to the four-fermion interaction, which is
marginal in the model at the classical level irrespective of the
values of the momentum exchange.} 
In terms of
the fermion propagator  $G^{(0)}_{\sigma} ({\bf k}, \omega )$ for each 
respective spin orientation, the vertex function to the given
perturbative order $\: i \Gamma^{(2)} ({\bf k}, \omega) \:$ reads
\begin{equation}
i \Gamma^{(2)} = - \frac{U^2}{(2 \pi)^3} 
\int_{-\infty}^{\infty} d\omega_q \int d^2 q \sum_{\sigma} 
G^{(0)}_{\sigma} ({\bf q} + {\bf k}, \omega_q + \omega )
G^{(0)}_{\sigma} ({\bf q}, \omega_q )
\label{vertex}
\end{equation}
The fermion propagator to be used in our model is
\begin{equation}
G^{(0)}_{\sigma} ({\bf q},\omega_q ) = \frac{1}{\omega_q -
 t (q^2_x - q^2_y) + i\epsilon \: \rm{sgn} \: \omega_q }
\end{equation}
which takes into account that the dispersion relation
cannot be linearized at the singularity. The most important
point in the computation of (\ref{vertex}) is that the virtual
states in the loop have to be kept within the band determined by
the cutoff, i.e. only states with energies between $-\Lambda $
and $\Lambda$ are allowed. This bandwith cutoff has the virtue
of preserving the SO(1,1) symmetry of the model, that is nothing
but the invariance under the continuous set of transformations
\begin{equation}
\left(  \begin{array}{c}
         q_x'  \\   q_y'   
       \end{array}   \right)  = 
 \left(  \begin{array}{cc} 
     \cosh \: \phi  &  \sinh \: \phi   \\
     \sinh \: \phi  &  \cosh \: \phi 
       \end{array}    \right) 
\left(  \begin{array}{c}
         q_x   \\  q_y  
       \end{array}   \right) 
\label{chv}
\end{equation}
The change of variables (\ref{chv}) preserves the form of the
dispersion relation (\ref{disp}) and it is an exact symmetry 
of the quantum theory. The implementation of a cutoff that does
not break the SO(1,1) invariance of the model is an important
ingredient of the calculation since, as we will see, it enables us
to reconstruct the complete expression of $\: i \Gamma^{(2)} ({\bf
k}, \omega ) \:$ from just the dependence on, say, $k_x$ and
$\omega$. 

The loop integral in (\ref{vertex}) is then most easily
performed by making a change of variables that introduces the
lines of constant energy. In the sector ${\cal R}_1 \equiv
\{ {\bf q} \mid q_y \geq \left| q_x \right| \}$, for instance,
we introduce the variables $r$ and $\phi$ by
\begin{eqnarray}
 q_x  & = &  r \; \sinh \: \phi  \nonumber   \\
 q_y  & = &  r \; \cosh \: \phi   
\end{eqnarray}
The region of integration in that sector is mapped to $-\infty <
\phi < +\infty$, $0 < r < \Lambda $. By doing first the
integration in the $\omega_q $ complex plane, we get a partial
contribution $\: i \Gamma^{(2)}_{{\cal R}_1} \:$ to (\ref{vertex}) 
\begin{eqnarray}
\lefteqn{i \Gamma^{(2)}_{{\cal R}_1} = }   \nonumber \\ 
 &  =  &   -i \frac{U^2}{2 \pi^2} 
\int_{0}^{\sqrt{\Lambda /t}} dr \int_{\alpha }^{\infty} d\phi
\: r \frac{1}{\omega - t k^2_x - 2tk_x r \: \sinh \: \phi
+ i \epsilon + i \epsilon \: \rm{sgn} \: (\omega - tr^2 )}
\label{inte}
\end{eqnarray}
where $\alpha = {\rm arcsh}((tr^2 - tk^2_x)/(2tk_x r)) $ 
and we are computing setting
$\omega > 0$, $k_x > 0$ and $k_y = 0$. As we have mentioned
before, the complete dependence of the vertex function on ${\bf
k}$ and $\omega $ can be reconstructed, in general, by using the
SO(1,1) symmetry of the model, as long as the vertex can only
depend on invariants under that transformation group.

The angular integral in (\ref{inte}) can be easily done, while
the remaining integral in the $r$ variable cannot be directly
obtained. It is much simpler, however, to extract its $\Lambda $
dependence, which turns out to be
\begin{equation}
i \Gamma^{(2)}_{{\cal R}_1} \approx i \frac{1}{8 \pi^2}
 \frac{U^2}{t} \; \log \: \Lambda
\end{equation}
It can be checked that there is a similar contribution from the
sector $q_y \leq - \left| q_x \right| $ and that the total
contribution of the region with $q^2_x - q^2_y > 0$ equals that
of the region $q^2_y - q^2_x > 0$. We have therefore
\begin{equation}
i \Gamma^{(2)} \approx i \frac{1}{2 \pi^2} \frac{U^2}{t} \; 
\log \: \Lambda
\label{cutoff}
\end{equation}
The presence of this dependence of the vertex function on the
cutoff is a key result for the model, since it implies that the
bare four-fermion interaction is renormalized as the
bandwith cutoff is reduced towards the Fermi level. This
property establishes a clear difference with the behavior of the
forward scattering
coupling constant in Fermi liquid theory, where it is shown that
it remains marginal to all orders in perturbation theory. In our
case the cutoff dependence of (\ref{cutoff}) can be easily
understood from the physical interpretation of the diagram in
Fig. 2. The logarithm of $\Lambda $ in (\ref{cutoff}) is
nothing but a measure of the number of particle-hole excitations
in which a particle below the Fermi level is promoted to an
empty state above it. The logarithmic divergence can be
alternatively seen as the divergence in the count of
particle-hole transitions when the momentum transfer accross the
Fermi line is made arbitrarily small.

The complete dependence of the vertex function on $k_x$ and
$\omega $ can be obtained starting from the knowledge of its
imaginary part. This arises from the sum of four different
contributions like that in (\ref{inte}). We get
\begin{equation}
{\rm Im} \: \Gamma^{(2)} = \frac{1}{4 \pi} \frac{U^2}{t} \left(
\frac{ \left| \omega + tk^2_x \right|}{ tk^2_x} -
\frac{ \left| \omega - tk^2_x \right|}{ tk^2_x} \right)
\end{equation}
Given that this object can only depend on quantities invariant 
under the transformations (\ref{chv}), we may reconstruct the
full expression
\begin{equation}
{\rm Im} \: \Gamma^{(2)} ({\bf k}, \omega) = 
 \frac{1}{4 \pi} \frac{U^2}{t} \left(
\frac{ \left| \omega + \varepsilon ({\bf k}) \right|}
 { \varepsilon ({\bf k}) } -
\frac{ \left| \omega - \varepsilon ({\bf k}) \right|}
 { \varepsilon ({\bf k})} \right)
\label{im}
\end{equation}
Moreover, we know that $\Gamma ({\bf k}, \omega)$ must have an
analytic continuation from the half-line $\omega > 0$ into the
upper half-plane of the complex variable $\omega $ \cite{landau}. 
This fact,
together with the knowledge of the imaginary part (\ref{im}),
would suffice to determine the real part of the vertex
function. Alternatively, one may introduce a different
implementation of the cutoff, which makes the real part of
$\Gamma^{(2)}$ directly available. The second version of the
computation is sketched in Appendix I. There we apply the
bandwith cutoff to the frequency integral in (\ref{vertex}),
letting the remaining integrals in $q_x$ and $q_y$ from
$-\infty$ to $+\infty$. Quite amazingly, these can be done
analytically by passing to the complex plane, producing the same
imaginary part as quoted in (\ref{im}) and the real part
\begin{equation}
{\rm Re} \: \Gamma^{(2)} ({\bf k}, \omega) = \frac{1}{2 \pi^2}
\frac{U^2}{t} \; \log \: \frac{\left| \varepsilon ({\bf k})
\right| \Lambda}{\left| \omega^2 - \varepsilon ({\bf k})^2
 \right|}  +  \frac{1}{2 \pi^2} \frac{U^2}{t} \frac{\omega}
 {\varepsilon ({\bf k}) } \; \log \: \left| \frac{\omega -
 \varepsilon ({\bf k}) }{\omega + \varepsilon ({\bf k}) }
 \right|  
\label{re}
\end{equation}
It can be easily seen that the expression (\ref{im}) corresponds
for $\omega > 0$ to the imaginary part of the analytic function
built from (\ref{re}), with definite choices of the branch cuts
for the logarithms. This shows the agreement between the two
different implementations of the bandwith cutoff and finishes up
the complete determination of the vertex function to the
one-loop order.

The renormalization of the vertex function under variations of
the cutoff $\Lambda $ has important consequences from the
physical point of view. Within the path integral formulation
defined by (\ref{pi}), we may interpret $\Gamma^{(2)}$ as a
quantum correction to the four-fermion interaction term. We 
define therefore the four-fermion effective interaction by
\begin{equation}
-i V_{eff} = -i U + i \frac{1}{2 \pi^2}
\frac{U^2}{t} \; \left(  \log \: 
   \frac{\left| \varepsilon ({\bf k})
\right| \Lambda}{\left| \omega^2 - \varepsilon ({\bf k})^2
 \right|}  +  \frac{\omega}
 {\varepsilon ({\bf k}) } \; \log \: \left| \frac{\omega -
 \varepsilon ({\bf k}) }{\omega + \varepsilon ({\bf k}) }
 \right|  \right)
\label{ueff}
\end{equation}
The fundamental observation now is that, if $V_{eff}$ is to play
the role of effective interaction with which observables are
computed in the quantum theory, it cannot depend on a
particular choice of the cutoff $\Lambda $. This means that, as
we reduce for instance the value of $\Lambda $, the value of the
bare coupling constant $U$ has to be adjusted so that the
effective interaction of the theory remains cutoff independent.
We recover in this way the idea underlying the original version
of the renormalization group, in that the hamiltonian of the
theory moves in the space of couplings as the 
length scale is varied\cite{wil}. 
By impossing the cutoff independence of
$V_{eff}$ 
\begin{equation}
\Lambda \frac{d}{d \Lambda} V_{eff} \equiv 0
\end{equation}
we get from (\ref{ueff}), to second order in perturbation
theory,  
\begin{equation}
\Lambda \frac{d}{d \Lambda} U(\Lambda ) = \frac{1}{2 \pi^2}
\frac{U^2}{t}
\label{ul}
\end{equation}
This equation governs the flow of the coupling constant as the
cutoff is lowered towards the Fermi level, leading to an
effective theory for the low-energy processes of the
model. We see that, in the case of a repulsive interaction we
are interested in ($U > 0$), the bare four-fermion coupling
constant decreases as $\Lambda \rightarrow 0$. It would seem,
therefore, that perturbation theory should become more reliable
at low energies if the starting value of $U(\Lambda )$ was
already small. One has to bear in mind, however, that the
perturbative expansion in the model is given in powers of $U/t$,
so that one has to check if the ``hopping'' parameter $t$ bears
any dependence on $\Lambda $ to ascertain whether the flow at
low energies is to weak coupling or to strong coupling in terms
of $U/t$. We consider the renormalization of $t$ in the next
section, in the context of the quantum corrections to the
two-point function.

\section{Self-energy corrections}

We undertake at this point the analysis of the self-energy
function of the model, with the aim of studying the
renormalization of the bare dispersion relation (\ref{disp}).
The self-energy $\Sigma ( {\bf k}, \omega )$ is given by the
irreducible part of the corrections to the two-point function
and it is related to the full propagator by the 
equation\cite{landau}
\begin{equation}
\frac{1}{G} = \frac{1}{G^{(0)}} - \Sigma
\end{equation}
The first perturbative contribution to $\Sigma $ is shown
diagrammatically in Fig. 3(a), where the dashed line represents
the interaction potential $V({\bf k}, \omega)$. We have started
in our model with a constant interaction $V({\bf k})$, 
and for this reason it is clear that the mentioned diagram
cannot give rise to a dependence on the
external frequency or momentum. Under these conditions the
one-loop diagram may only produce a renormalization of the
two-point function by a constant, i.e. a renormalization of the
chemical potential. We regard, however, this no-renormalization
of the rest of parameters as a mere accident, since any slight
modification of the potential (that still may keep it
short-ranged) already introduces some nontrivial dependence
$V({\bf k}, \omega)$. We will recall this observation later on,
when discussing the renormalization of $\varepsilon ({\bf k})$.

The first nontrivial contributions to the self-energy in our
model are given by the two-loop diagrams shown in Figs. 3(b) 
and 3(c). Since
we are dealing with a constant interaction the dashed lines can be
contracted to a point, making the two diagrams look the
same. The only difference between diagram 3(b) and diagram 3(c)
is that the first comes with a relative factor of $-2$ with
respect to the second due to the propagation of the two spin
orientations around the fermion loop. The net effect is taken
into account by computing diagram 3(b) for one of the spin
orientations. We implement the same regularization adopted
before, preserving the SO(1,1) invariance of the model. Thus we
may take advantage of the one-loop particle-hole polarizability
determined in the previous section. That is, at the two-loop
level we have
\begin{equation}
i \Sigma^{(2)} ({\bf p}, \omega_p ) = 
\frac{U^2}{(2 \pi)^3} \int d
\omega_k \: d^2 k \; G^{(0)} ({\bf p} - {\bf k}, \omega_p -
\omega_k ) \Gamma^{(2)} ({\bf k}, \omega_k )
\end{equation}
The additional loop integral is taken within the region
$-\Lambda < \varepsilon ({\bf k}) < \Lambda$. By dimensional
reasons, it is clear that for $\omega_p ,\varepsilon ({\bf p})
<< \Lambda$ the dominant behavior is given by the terms
proportional to $\omega_p $ and
$\varepsilon ({\bf p})$, as long as the cutoff is the only
explicit scale in the theory and these quantities saturate the
dimensions of the inverse propagator. Those are, on the other
hand, the contributions in which we are interested in, since the
term proportional to $\omega_p $ in the self-energy renormalizes
the scale of the electron field operator while the term
proportional to $\varepsilon ({\bf p})$ may give rise to a
renormalization of the ``hopping'' parameter $t$.

For the sake of a practical computation of $\Sigma ({\bf p},
\omega_p)$, we perform the loop integral by making a change of
variables of the type (\ref{chv}) in the respective sectors of
positive and negative energy. In all cases, we first carry out the
integral over the hyperbolic angle $\phi $, which is finite
provided we keep the external momentum ${\bf p}$ different from
zero. To obtain the term linear in $\omega_p$, we
take the derivative with respect to $\omega_p$ and set $\omega_p
= 0$. In the region $k_y \geq \left| k_x \right|$, for instance,
we get
\begin{eqnarray}
\left. \frac{\partial }{\partial \omega_p} {\rm Re} \:
\Sigma^{(2)}_{{\cal R}_1} \right|_{\omega_p = 0} & = & 
 \frac{1}{16 \pi^4} \frac{U^2}{t^2} \int^{\Lambda} dr \:
\frac{1}{r} \int_{0}^{\infty} dx \sum_{n = \pm 1} 
 \frac{x - n(1-\epsilon^2)}{\sqrt{\left( (x - n(1-\epsilon^2))^2
+ 4\epsilon^2 \right)^3 } }    \nonumber       \\
  &  &   \times \;    \left( -\log \: \left|x^2 - 1\right|
+ x \: \log \: \left| \frac{x-1}{x+1} \right| \right) \nonumber \\
  &  & + \frac{1}{8 \pi^4} \frac{U^2}{t^2} \int^{\Lambda} dr \:
\frac{1}{r} \int_{0}^{1} dx \sum_{n = \pm 1} 
 \frac{1}{\sqrt{ (x - n(1-\epsilon^2))^2
+ 4\epsilon^2  } }          \nonumber     \\
  &  &  \times \;      \log \: \frac{2 \epsilon}{x -
 n(1-\epsilon^2) +   \sqrt{ (x - n(1-\epsilon^2))^2
+ 4\epsilon^2  } }  
\label{sig1}
\end{eqnarray}
where we use the parameter $\epsilon \equiv p_x/r$ as an
infrared regulator (we have set for simplicity $p_y = 0$). The
first term arises from the real part of $\Gamma^{(2)}$ times the
imaginary part of $G^{(0)}$ while the second arises from
the reverse combination. The terms containing the
explicit dependence on the cutoff
from $\Gamma^{(2)}$ cancel out in $\Sigma^{(2)}$, what is
consistent with the fact that they are a local insertion
in the one-loop diagram of Fig. 3(a). Another remarkable fact
regarding equation (\ref{sig1}) is that the infrared regulator
$\epsilon $ cannot be sent to zero, if we want to maintain
finite the above expression. This will have important
consequences, as it turns out that the dependence of
$\Sigma^{(2)}$ on the cutoff becomes entangled, in principle,
with a dependence on the external momentum.

Putting together the pieces coming from the different
integration regions, we arrive at the form of the linear 
dependence of $\Sigma^{(2)}$ on $\omega_p$
\begin{eqnarray}
\lefteqn{ {\rm Re}\: \Sigma^{(2)} ({\bf p} \rightarrow 0, \omega_p) 
     \approx  }         \nonumber        \\
 & \approx  & \omega_p \frac{1}{4 \pi^4} \frac{U^2}{t^2} 
\int^{\Lambda} dr \:
\frac{1}{r} \left(2 \; \log \: 2 \; \log \: \left| \frac{
\varepsilon ({\bf p}) }{t r^2} \right| + c_1 \right) 
                            \nonumber   \\
 &  \approx  & \omega_p \frac{1}{4 \pi^4} \frac{U^2}{t^2} \left(
 -2 \; \log \: 2 \; \log^2 \: \left( \Lambda / \left| 
p^2_x - p^2_y \right|^{1/2}  \right) + c_1 \; \log \: \Lambda
  \right) 
\label{res}
\end{eqnarray}
with $c_1 \approx -5.896 $.
We have used again the SO(1,1) symmetry to reconstruct the
dispersion relation from the momentum dependence of the
$\epsilon$ parameter. The above expression shows that there is a
renormalization of the scale of the electron field as the cutoff
is varied, though at this point this effect cannot be assessed
in a way independent of the electron state. In other words, to
the coefficient $c_1$ of the $\: \log \: \Lambda \:$ term in
(\ref{res}) we should add the $\: \log \: \left| 
p^2_x - p^2_y \right|^{1/2}$ coming from the squared logarithm
contribution, leading to the result of a wavefunction
renormalization depending on the electron momentum. This is
what, in the field theory language, is called a nonlocal
divergence, since it cannot be removed in the framework of 
quantum field theory by the subtraction of terms with power-law
dependence on the frequency or the momentum. In the present
situation it may also appear doubtful that the $\: \log \: \left| 
p^2_x - p^2_y \right|^{1/2} \; \log \: \Lambda \:$ cutoff dependence
can be removed by a convenient redefinition of the couplings in
the theory. However, the main difference of the condensed matter
framework with respect to quantum field theory is that one may
envisage the renormalization of the interaction potential and
the Fermi surface, which are, in general, functions of the
electron momentum. This approach to the problem of interacting
fermions is not usually adopted, mainly because in Fermi liquid
theory the interaction, taken as a manifold of couplings, is not
irrelevant only in a very reduced number of channels. We have
seen, though, that this type of restriction does not apply to
our model, as the four-fermion interaction remains marginal no
matter what is the kinematics of the momenta involved on it.
The fact that the first nontrivial contribution in perturbation
theory introduces a leading term of the type $\log^2 \: \Lambda$,
instead of a simple logarithm, points also at the solution of
the problem. The absence of a $\: \log \: \Lambda \:$ contribution to
first order in perturbation theory is just a consequence of
having fine-tunned the bare potential $V({\bf k})$ to unity,
which now appears clearly unnatural since any slight 
(scale independent) deviation from that form introduces
significant renormalization effects. We will follow this line of
thought in order to apply with success the renormalization
program in the next section.

We complete the discussion of the self-energy corrections by
considering the contributions proportional to the dispersion
relation $\varepsilon ({\bf p})$. Their computation is carried
out by differenciating now $\Sigma^{(2)}$ with respect to
$p^2_x$ or $p^2_y$, and then setting them to zero except where
they act as an infrared regulator. Details of the calculation
are given in Appendix II. We finally arrive at the result
\begin{eqnarray}
\lefteqn{ {\rm Re} \Sigma^{(2)} ({\bf p} \rightarrow 0, 
  \omega_p = 0)   \approx  }   \nonumber   \\
  & \approx  &   (p^2_x - p^2_y)
\frac{1}{4 \pi^4} \frac{U^2}{t} \int^{\Lambda} dr \:
\frac{1}{r} \left((-2 \; \log \: 2 + \frac{3}{2})
 \; \log \: \left| \frac{
\varepsilon ({\bf p}) }{t r^2} \right| + c_2 \right) 
                            \nonumber   \\
 &  \approx  &  \varepsilon ({\bf p})
 \frac{1}{4 \pi^4} \frac{U^2}{t^2} \left(
 (2 \; \log \: 2 - \frac{3}{2})
 \; \log^2 \: \left( \Lambda / \left| 
p^2_x - p^2_y \right|^{1/2}  \right) + c_2 \; \log \: \Lambda
  \right) 
\label{resp}
\end{eqnarray}
with $c_2 \approx 3.839$. We can apply to this expression the
same remarks made for the contributions linear in $\omega_p$ to
the self-energy. It is worth noting that the dependence on the
cutoff in (\ref{resp}) is not the same as that renormalizing the
scale of the electron field in equation (\ref{res}). This means
that, besides the wavefunction renormalization also present in
(\ref{resp}), there is a left-over that leads directly to the
modification of the dispersion relation. However, the nonlocal
character of the $\Lambda$-dependence prevents again to
understand that correction, as it stands, as a pure
renormalization of the hopping parameter $t$. This issue will be
conveniently clarified in the next section.

\section{Renormalization}

The goal of applying the renormalization group program to a
statistical field theory of the type we are studying is to
remove all the dependence on the cutoff from observable quantities
by an appropriate redefinition of the couplings (in 
general sense) of the theory. This philosophy corresponds to
the point of view of relating different cutoff-dependent bare
theories to a unique cutoff-independent renormalized theory\cite{zj,amit}. 
In quantum field theory there are theorems that establish, in
general, whether a given model is renormalizable, so that all
the divergences as $\Lambda \rightarrow \infty$ can be absorbed
in the parameters of the theory. In the context of condensed
matter systems, however, the criteria for interacting fermion
systems are more unclear, as the renormalization group approach
has been developed in recent years and limited mainly to known
paradigms, like that of Fermi liquid theory, and to models with 
properties common to quantum field theories. It turns out,
though, that the number of degrees of freedom that are available
in the renormalization of a statistical fermion system is much
larger than that in a quantum field theory. In the latter the
number of marginal couplings is always finite (except in the
case of spatial dimension equal to one) while in a fermion
system with a nonvanishing charge density the Fermi surface (or
line) and the interaction potential should be taken as dynamical
variables influencing one to each other. In other words, there
are two different functions, not just parameters, that can be
set to enable the cutoff independence of the renormalized
theory. The situation is in that respect reminiscent of 
string theory with massless background fields. Although the
string interaction is clearly nonlocal, there is always a choice
of the background condensates to each perturbative order that
keeps the theory conformal invariant\cite{str}.

Coming back to our model, we should allow for redefinitions of
the dispersion relation $\varepsilon ({\bf k})$ and the
interaction potential $V({\bf k})$ in order to get rid of the
nonlocal cutoff dependences of the self-energy in (\ref{res})
and (\ref{resp}). On the one hand, the saddle point in 
$\varepsilon ({\bf k})$ has a topological character, in the
sense that it cannot be removed by any smooth deformation of the
dispersion relation. Therefore, we concentrate on the
possibility of a modification of the bare four-fermion
interaction by the presence of the van Hove singularity. There
are two different ways in which this can be understood, namely
from a physical and a technical point of view. On physical
grounds, the fact that the renormalization coefficients in
(\ref{res}) and (\ref{resp}) diverge as ${\bf p} \rightarrow 0$
means that the renormalization process is out of control as one
approaches the singularity. This instability can be interpreted
as a signal that the physics is changing drastically at the flat
region near the origin, in a way that our initial bare theory
cannot encompass. On technical grounds, the instability can be
traced back to the significant change that the quantum corrections
bring about in the interaction potential at very low energies.
This can be appreciated in the one-loop polarization tensor 
(\ref{re}), which corrects the classical four-fermion
interaction. The two effective interactions in the
right-hand-side of (\ref{re}) are the very source of the squared
logarithm cutoff-dependence in the self-energy function, and
they give the hint that similar terms have to be included in the
bare interaction to absorb the nonlocal quantum corrections.
From the point of view of a naive scaling, a logarithmic
correction to the bare potential does not change the marginal
character of the interaction, though it modifies in general
the real space potential at long distances by a $1/r^2$ tail.

There is a sensible difference between the effective
interactions that appear in (\ref{ueff}), namely
\begin{eqnarray}
V_1 & \sim & \; \log \: \left| \omega^2 - \varepsilon ({\bf k})^2
  \right| + i A_1        \nonumber     \\
V_2 & \sim & \frac{\omega}
 {\varepsilon ({\bf k}) } \; \log \: \left| \frac{\omega -
 \varepsilon ({\bf k}) }{\omega + \varepsilon ({\bf k}) }
 \right|  + i  A_2
\end{eqnarray}
and those that can enter in a physical potential. A retarded
interaction in real space gives rise to a potential in
$(\omega, {\bf k})$ space that has an analytic extension from
the real line to the upper half-plane of complex frequencies. As
is well-known, though, the effective potentials that arise from
the particle-hole polarizability $\Gamma^{(2)}$ are analytic up
to $\omega = 0$ \cite{landau}. 
The imaginary parts of $V_1$ and $V_2$ are even
functions of $\omega$, instead of odd functions as it should be
to combine with the real part of respective analytic functions.
For this reason, even if we renormalize the bare potential to
absorb the nonlocal $\Lambda$-dependences that appear in the
self-energy corrections, the effect of this renormalization can
only cancel part of the effective interaction.

The other obvious condition that is required to any admissible
modification of the bare potential is to have a real part that is
an even function of $\omega $, leading to a real
interaction in real space. Thus, we arrive at the conclusion
that the real parts of the effective potentials $V_1$ and $V_2$
produced by the particle-hole polarizability are just the only
dimensionless expressions that can be thought of. The  second
potential $V_2$ generates however too singular $\: \log^3 \:
\Lambda \:$ contributions when inserted in the one-loop self-energy
diagram, that cancel only for the specific imaginary part
arising from $\Gamma^{(2)}$ (and did not show up therefore in 
$\Sigma^{(2)}$). It seems then that we are only left with the
real part of $V_1$ in our purpose of building up an analytic
potential that may remove the nonlocal $\Lambda$-dependences of
$\Sigma^{(2)}$. There are actually two possibilities, depending
on whether the argument of the logarithm is scaled with
$\varepsilon ({\bf k})^2$ or with $\omega^2$. We perform
therefore a renormalization of the potential in the form
\begin{equation}
U \tilde{V} ({\bf k}, \omega) = U V + 
  \frac{U^2}{t} V^2 \left( a_1 \; \log \: 
\left(  \frac{ \omega^2 - \varepsilon ({\bf k})^2}
   {\varepsilon ({\bf k})^2} \right)
 + a_2 \; \log \: \left( \frac{ \omega^2 - \varepsilon ({\bf k})^2}
  {\omega^2 } \right)  \right) + O \left( U (U/t)^2 \right)
\label{ret}
\end{equation}
and demand that $a_1$ and $a_2$ are such that, by insertion of
$\tilde{V}({\bf k}, \omega)$ in the one-loop self-energy diagram, the
nonlocal $\Lambda$-dependences cancel up to the two-loop order.
$a_1$ and $a_2$ are then uniquely determined, since as we
already remarked there are independent quantum corrections to
the scale of the electron field and to the dispersion relation
$\varepsilon ({\bf k})$. 

When used at the one-loop level, the
potential (\ref{ret}) gives to order $U^2/t^2$ a contribution to
the self-energy
\begin{eqnarray}
{\rm Re} \: \Sigma^{(1)} & \approx & \omega_p 
   \frac{1}{8 \pi^4} \frac{U^2}{t^2} \left(
 -\frac{a_1}{2} \; \log^2 \: \left( \Lambda / \left| 
p^2_x - p^2_y \right|^{1/2}  \right) - 2.193 \: a_1 \; \log \:
   \Lambda  \right.  \nonumber     \\
 &  & \;\;\;\;\;\;\; + \left. \frac{3}{2} a_2 \; \log^2 \: \left( 
    \Lambda / \left| 
p^2_x - p^2_y \right|^{1/2}  \right) - 2.193 \: a_2 \; \log \:
   \Lambda  \right)     \nonumber   \\
 &  & +  \varepsilon ({\bf p}) 
   \frac{1}{8 \pi^4} \frac{U^2}{t^2} \left(
 \frac{a_1}{4} \; \log^2 \: \left( \Lambda / \left| 
p^2_x - p^2_y \right|^{1/2}  \right) + 1.847 \: a_1 \; \log \:
   \Lambda  \right.    \nonumber     \\
 &  & \;\;\;\;\;\;\; + \left. \frac{a_2}{4} \; \log^2 \: \left( 
    \Lambda / \left| 
p^2_x - p^2_y \right|^{1/2}  \right) - 2.153 \: a_2 \; \log \:
   \Lambda  \right)  
\end{eqnarray}
If we now set $a_1 = -14 \; \log \: 2 + 9$ and $a_2 = -2 \; \log \: 2
 + 3$, we see that all the nonlocal cutoff-dependences are
removed from the self-energy function to order $U^2/t^2$
\begin{eqnarray}
\Sigma   & = & \Sigma^{(1)} + \Sigma^{(2)} + \ldots 
                   \nonumber    \\
  & \approx &  \omega_p \frac{1}{4 \pi^4} \frac{U^2}{t^2}
 \left( -6.893 \; \log \: \Lambda \right) 
  - \varepsilon ({\bf p}) \frac{1}{4 \pi^4} \frac{U^2}{t^2}
 \left( -1.452 \; \log \: \Lambda \right) + O \left( (U/t)^3
   \right) 
\label{self}
\end{eqnarray}
This amounts to accomplish half the renormalization program in
the model, since there still remain the local
$\Lambda$-dependences that have to be removed by the usual
procedure of absorbing them in the scale of the bare electron
field and in the bare hopping parameter.

We require the cutoff-independence of the full Green function 
given by
\begin{eqnarray}
\frac{1}{G} & = & \frac{1}{G^{(0)}} - \Sigma  \nonumber  \\
  & \approx & Z^{-1}_{\Psi}(\Lambda ) \left( \omega_p - t(\Lambda)
(p^2_x - p^2_y) \right)   \nonumber  \\
 &  & + Z^{-1}_{\Psi}(\Lambda ) \left(\omega_p - \varepsilon ({\bf p})
 \right)   \frac{1}{4 \pi^4} \frac{U^2}{t^2}
 \left( 6.893 \; \log \: \Lambda \right)  \nonumber  \\
 &  & + Z^{-1}_{\Psi}(\Lambda ) \varepsilon ({\bf p}) \frac{1}{4 \pi^4} 
    \frac{U^2}{t^2}
 \left( 5.441 \; \log \: \Lambda \right) + O \left( (U/t)^3
   \right) 
\label{invp}
\end{eqnarray}
since this object leads to observable quantities in the quantum
theory. In (\ref{invp}) $Z^{1/2}_{\Psi}(\Lambda)$ represents the scale
of the bare electron field compared to that of the
cutoff-independent electron field
\begin{equation}
\Psi_{bare} (\Lambda ) = Z^{1/2}_{\Psi}(\Lambda) \Psi
\end{equation}
From 
\begin{equation}
\Lambda \frac{d }{d \Lambda} G^{-1}  \equiv 0
\end{equation}
we get the differential equations, to the two-loop order,
\begin{eqnarray}
 \Lambda \frac{d }{d \Lambda} \; \log \: Z_{\Psi}(\Lambda) 
  & = & 6.893 \frac{1}{4 \pi^4} \frac{U^2}{t^2} \label{zl} \\
 \Lambda \frac{d }{d \Lambda} \; t(\Lambda) 
  & = & 5.441 \frac{1}{4 \pi^4} \frac{U^2}{t} \label{tl}
\end{eqnarray}
The quantities $Z_{\Psi}(\Lambda) $ and $t(\Lambda) $ play the
role of effective parameters that reflect the behavior of the
quantum theory as $\Lambda \rightarrow 0$ and more states are
integrated out from high-energy shells of the dispersion
relation. The equations (\ref{zl}) and (\ref{tl}) have to be
solved in conjunction with equation (\ref{ul}) for $U(\Lambda)$,
that completes the set of coupled renormalization group
equations. 

At last, we have accomplished successfully our purpose of
absorbing all the dependences on the cutoff by renormalizing the
interaction and other parameters of the theory. In the process
we have been able to interpret the nonlocal $\Lambda$-dependent
quantum corrections, that obstructed the knowledge about the
physics close to the singularity, as a renormalization of the
interaction potential. This may appear a little bit surprising,
since it seems that we had gained a complete predictability about
the behavior of the theory at arbitrarily small energies.
However, this may not be necessarily the case. The potentials 
in (\ref{ret}) that we have introduced to renormalize the
interaction grow large at very small values of $\omega^2$ or
$\varepsilon ({\bf k})^2$. Even if our renormalization group
approach predicts that in certain regime the effective coupling
constant $U(\Lambda)/t(\Lambda)$ becomes small, so that
perturbation theory may become reliable, there cannot be an
absolute statement about the weakness of the interaction, since
at very small frequencies or very large distances the
interaction (\ref{ret}) enters necessarily the nonperturbative
regime. It seems, therefore, that we have not obviated yet the
existence of some limitation in the close approach to the
singularity, at least while we do not get information about the
form of the successive renormalizations of $V({\bf k}, \omega)$
to higher orders in perturbation theory.

\section{Physical properties}

The solution of the coupled set of renormalization group
equations (\ref{ul}), (\ref{zl}) and (\ref{tl}) gives, in the
limit $\Lambda \rightarrow 0$, the behavior of the statistical
field theory at low energies. In principle, the 
nontrivial renormalizations of the wavefunction of the electron,
the hopping parameter and the four-fermion coupling constant are
quite relevant results, since none of them have a counterpart in
Fermi liquid theory\cite{sh,pol}. 
The flow of the scale of the electron field
in (\ref{zl}), for instance, is such that it reflects the
suppression of the weight of fermion quasiparticles at the Fermi
level. This agrees with the conclusion obtained in the context
of a purely phenomenological model of the copper-oxide
superconductors\cite{varma}. 
The flow of the hopping parameter in (\ref{tl})
gives information about how the dispersion relation is
renormalized in the low-energy limit. From that equation we may
already draw the conclusion that $\varepsilon ({\bf k})$ has in 
any event a tendency to become flatter near the singularity.
This behavior is also
in agreement with experimental observations\cite{photo,abr}.

The main issue, however, is the determination of the coupling
regime to which the model flows at low energies. This is not
just given by the flow of the four-fermion coupling constant
$U$, since the parameter that enters in the power series
expansion of perturbation theory is $U/t$. Thus, although
$U(\Lambda)$ decreases as $\Lambda \rightarrow 0$, $t(\Lambda)$
also does and the consideration of the behavior of $U/t$
requires further detail.

From (\ref{ul}) and (\ref{tl}) we obtain\footnote{It may seem
that in (\ref{flow}) there is a $(U/t)^3$ term missing from
coupling constant renormalization, but this contribution belongs
to the RPA sum already encoded in the term at the
right-hand-side of (\ref{ul}).}
\begin{equation}
\Lambda \frac{d}{d \Lambda} \frac{U}{t} = \frac{1}{2 \pi^2}
 \left( \frac{U}{t} \right)^2 - \frac{5.441}{4 \pi^4} \left(
\frac{U}{t} \right)^3  
\label{flow}
\end{equation}
We regard the expression at the right-hand-side of (\ref{flow})
as the beta function of the effective coupling $U/t$, whose
zeros correspond to fixed points of the flow (see Fig. 4). 
Apart from the fixed point at the origin, another fixed point
shows up at a value of the coupling $(U/t)_{critical} \approx 2
\pi^2/5.441 $. One may question whether this point is a sensible
feature of the model, as it may not lie in the region in which
perturbation theory can be trusted. We think that it is realistic
to consider that the perturbative expansion is more precisely a
power series in the parameter $U/(2 \pi^2 t)$, taking into
account the phase space factors. Thus, it is very likely that
the nontrivial fixed point that we obtain within perturbation
theory is the evidence for a fixed point that remains at the
nonperturbative level. A plot of the beta function $\beta (U/t)$
in the right-hand-side of (\ref{flow}) is given in Fig. 4. The
points with $\beta > 0$ have a flow that leads towards the
origin as $\Lambda \rightarrow 0$ while those with $\beta < 0$
flow towards increasing values of $U/t$, so that the nontrivial
fixed point at $(U/t)_{critical}$ is unstable in the infrared.
The solution of (\ref{flow}) can be given in closed implicit
form 
\begin{equation}
\frac{U}{t} = \frac{U_0/t_0}{1 - \frac{1}{2 \pi^2}
\frac{U_0}{t_0} \; \log \left(\frac{\Lambda}{\Lambda_0}\right)
+ \alpha \frac{U_0}{t_0} \; \log \left(\frac{U}{t}
\frac{t_0}{U_0} \right) - \alpha \frac{U_0}{t_0} \; \log 
\left| \frac{1 - \alpha U/t}{1 - \alpha U_0/t_0} \right| }
\label{sol}
\end{equation}
where $\alpha \approx 5.441/(2 \pi^2)$. It seems, therefore,
that the system has at least two different phases. One of them
is within the weak coupling regime, in which the theory becomes
asymptotically free in the low-energy limit. The other phase
goes into the strong coupling regime, leading presumably to
different physical properties that cannot be studied within
perturbation theory.

Next, the behavior of $t(\Lambda)$ can be easily understood in
the following way. From equations (\ref{ul}) and (\ref{tl}) one
may notice that the flow in $(t,U)$ space is given by straight
lines 
\begin{equation}
\frac{dU}{dt} = \frac{2\pi^2}{5.441}
\end{equation}
Then, it is clear that if we start with some initial value
$U_0/t_0 < 2\pi^2/5.441 $, the hopping parameter $t(\Lambda)$
will decrease to some finite value as $\Lambda \rightarrow 0$.
On the other hand, if $U_0/t_0 > 2\pi^2/5.441 $ the running
hopping $t(\Lambda)$ will have no limit in its approach to zero,
though our perturbative renormalization group methods become
unreliable before reaching that value. We observe anyhow a
definite trend of the hopping parameter towards decreasing
values in the process of renormalization, which may persist in
the strong coupling regime. This translates, in turn, into a
tendency of the dispersion relation to become flatter around the
singularity and, strictly speaking, a value of $t(\Lambda)
\approx 0$ would mean that the dispersion relation $\varepsilon
({\bf k}) = t(k^2_x - k^2_y)$ turns into a higher degree
singularity. Quite remarkably, the experimental studies of
copper-oxide superconductors by means of angle-resolved
photoemission spectroscopy show very flat regions of the
dispersion relation near the Fermi level\cite{photo,abr}. 
Similarly, an almost
dispersionless region arises from numerical studies of the
dynamics of a few holes in the half-filled two-dimensional
square lattice\cite{dagotto}. 
All this evidence points in the direction that
the correspondence with the physically relevant situation 
for the cuprates may take
place from the strong coupling phase of our model, where the
renormalization of the dispersion towards a flat band would
persist at the nonperturbative level.

Regarding the renormalization of the electron wavefunction, we
find again two different situations depending on whether we stay
in the weak or in the strong coupling phase. In the weak
coupling phase, we can integrate equation (\ref{zl}) in a
straightforward way taking the initial data in the interval
$\left( 0, (U/t)_{critical} \right)$
\begin{equation}
Z_{\Psi} (\Lambda ) = e^{6.893/(4\pi^4) \int^{\Lambda}
d\rho/ \rho \; U(\rho)^2 / t(\rho)^2 }
\end{equation}
As long as the running effective coupling
$U(\Lambda)/t(\Lambda)$ goes to zero when $\Lambda \rightarrow
0$ according to (\ref{sol}), we can only expect a very weak
reduction in the scale of the electron field at low
energies. Thus, the attenuation of quasiparticles should not 
differ much in this case from that in Fermi liquid theory. The
situation is completely different in the strong coupling phase,
since there the scaling dimension of the electron field is
actually changed from the noninteracting value. The anomalous
dimension is most easily computed in the particular instance in
which we sit at the critical coupling $(U/t)_{critical}$.
Applying now (\ref{zl}) we obtain 
\begin{equation}
Z_{\Psi} (\Lambda) = \Lambda^{(6.893 /(4\pi^4)) \left(
U /t \right)^2_{critical} }
\label{wfr}
\end{equation}
By renormalizing then the model to the Fermi level $\Lambda
\rightarrow 0$, the attenuation of the electron quasiparticles
becomes complete. This feature is similar to that proposed in
Ref. \cite{varma} in a phenomenological approach to the
copper-oxide superconductors. In our framework, however, it
acquires a more precise meaning. The renormalization effects are
so strong that they change at very low energies the form of the
electron propagator. The relation between the
unrenormalized and the renormalized
($\Lambda$-independent) electron Green function that follows
from (\ref{wfr}) 
\begin{equation}
  G({\bf k}, \omega)  = 
\Lambda^{(6.893 /(4\pi^4)) \left(
U /t \right)^2_{critical} } 
 \left. G({\bf k}, \omega) \right|_{\Lambda}
\end{equation}
leads to the scaling behavior
\begin{equation}
G(\rho {\bf k}, \rho^2 \omega) = \rho^{-2 + (6.893 /(2\pi^4)) 
\left( U /t \right)^2_{critical} } G( {\bf k}, \omega)
\label{scal}
\end{equation}
This anomalous scaling is another way of looking at the
breakdown of the quasiparticle picture. The equations
(\ref{wfr}) and (\ref{scal}) are only correct at the critical
point $(U/t)_{critical}$, that is an unstable fixed point.
Therefore it remains as an open question to what extent the
breakdown of the Fermi liquid picture persists when leaving the
critical point in the direction of the strong coupling phase.

We finish our discussion of the predictions that emerge from the
renormalization group approach by addressing the question of the
position of the Fermi level in the model. The results showing a
clear deviation from Fermi liquid behavior depend on the
coincidence of the Fermi level with the energy level of the
singularity, as it is the divergent density of states at that
point the source of the unconventional behavior. Actually, the
renormalization of the observables considered above still takes
place to a great extent provided that the energy difference
between the Fermi level and that of the singularity is much
smaller than any other energy scale in the system. It arises the
question, however, of whether this close approach of the two
levels is something natural, or rather any slight perturbation
may send the singularity away from the Fermi level. This problem
has been studied in Ref. \cite{us} from the point of view of a
system with variable number of particles and fixed chemical
potential. The conclusion reached there by a standard
renormalization group analysis is that as $\Lambda \rightarrow
0$ the running Fermi energy is captured by the singularity. This
feature is at odds with the unstable behavior of the Fermi
energy in Fermi liquid theory, where only a repulsive fixed
point is found for finite values of the chemical potential\cite{sh}
---this fact does not have however physical relevance,
as long as in closed systems the stability of the Fermi level is
ensured by the Luttinger theorem.

Rather than reproducing the analysis of Ref. \cite{us}, we give
here a simpler but more physical explanation of the pinning
mechanism. The argument is suited for a system that may receive
extra charge from a reservoir (as is the case of the
copper-oxide compounds) and does not have therefore fixed number
of particles. The number of particles in the reservoir is much
larger than that in the system, so that the flow of particles
into the latter is controlled by the chemical potential $\mu$ of
the reservoir. This quantity acts as a kind of external
``pressure'' of particles. However this external pressure is not
seen in all its magnitude in the two-dimensional system, since
it is partially counterbalanced there by the repulsion between the
electrons. The competition between the two effects is what sets
the value of the Fermi level $\varepsilon_{F}$ in the system. By
using a simple one-loop approximation we may express the
correction to the chemical potential as the frequency and
momentum-independent contribution of the Hartree diagram
\begin{equation}
\varepsilon_{F} \approx \mu + i \frac{U}{8\pi^3}\int d\omega \: d^2 k
G^{(0)} ({\bf k}, \omega)
\end{equation}
In terms of the density of states $n(\omega)$ we may write this
expression in the form
\begin{equation}
\varepsilon_{F} = \mu - \frac{U}{4\pi} \int^{\varepsilon_F}
d\omega \: n(\omega)
\label{selfc}
\end{equation}
We may obviate the influence of the particular shape of the
dispersion relation by differenciating (\ref{selfc}) with respect
to $\mu$, so that
\begin{equation}
\frac{d \varepsilon_{F}}{d \mu} = 1 -  \frac{U}{4\pi}
n(\varepsilon_F)  \frac{d \varepsilon_{F}}{d \mu}
\end{equation}
Then, we arrive at the differential equation
\begin{equation}
\frac{d \varepsilon_F}{d \mu} = \frac{1}{1 + \frac{U}{4 \pi}
n(\varepsilon_F ) }
\label{flev}
\end{equation}
which expresses how the filling level changes in the system
under variations of the external chemical potential. Equation
(\ref{flev}) shows that when the Fermi energy $\varepsilon_F$ is
very close to a level with a divergent density of states, as is
the case of the van Hove singularity, the Fermi level is very
weakly influenced by changes in the chemical potential of the
reservoir. Of course, this does not prove that the level of
filling has necessarily to fall close to the van Hove
singularity, but the simple argument outlined gives a notion of
the stability of the model with the particular filling
considered in this paper. On the other hand, with the methods
employed in Ref. \cite{us} it is possible to show that the
renormalization of the Fermi energy $\varepsilon_F$ keeps up
with the integration of the high-energy degrees of freedom. The
result leads to the dynamical effect of pinning the Fermi level
to the van Hove singularity as $\Lambda \rightarrow 0$. 
Quite remarkably, this feature is in 
correspondence with the experimental observation of many hole-doped
copper-oxide compounds, for which the Fermi level is found 
very close to a pronounced peak of the photoemission 
spectra\cite{photo}.

\section{Conclusions}

We have shown that a model of interacting electrons near a 2D
van Hove singularity can be treated by means of renormalization
group techniques. These have found recent application in the
general description of interacting fermion systems\cite{sh,pol}, 
which
require specific treatment given the distinctive character of
the many-body ground state, that is, the existence of an
extended Fermi surface. One of the main accomplishments of Refs.
\cite{sh,pol} has been the comprehension gained of Fermi liquid
theory and its perturbations. In fact, under very general
conditions like the isotropy of the Fermi surface and the
short-range character of the interaction, it appears that the
interacting fermion system falls necessarily into the Fermi
liquid universality class. The problem that we have addressed,
however, does not adhere to that well-established picture, as
the anisotropy of the dispersion relation demands an
appropriate treatment. As a result of that, we have found a
nontrivial renormalization group flow of the couplings in the
infrared. This flow is incompatible with conventional Fermi
liquid theory, making the model one of the few systems beyond
one dimension where the breakdown of a Fermi liquid is
explicitly manifested.

We find that, for sufficiently weak coupling, the interaction
strength is renormalized to zero, which is an attractive
fixed-point. This behavior is easy to understand, as it
corresponds to the standard picture in which the repulsive
interaction is screened at long distances by the particle-hole
polarizability of the medium. On the other hand, our results
suggest the existence of another unstable fixed-point, and of a
strong coupling phase beyond it. This new phase arises as a
consequence of the tendency of the dispersion relation to become
flatter near the van Hove singularity. As long as the hopping
parameter $t$ is renormalized towards zero in the infrared, the
effective coupling constant $U/t$ flows to larger values in the
new phase. The reduction of the dispersion of the band near the
singularity is in remarkable correspondence with the recent
experimental results from photoemission studies\cite{photo,abr}. 
Therefore,
there are solid reasons to believe that the new phase predicted
in the renormalization group approach should be a sensible
feature of the system. The precise knowledge of the nature of
the strong coupling phase would require, though, different
methods to that employed in this paper. Preliminary studies of
finite clusters with next-to-nearest-neighbor interactions
(excluding therefore a spin density wave instability) seem to
indicate a trend towards ferromagnetic order\cite{prel}, 
what is consistent
with the development of a very flat band.

The attenuation of the electron quasiparticles that we have
found in our model is in agreement with a
phenomenological approach to the copper-oxide 
superconductors\cite{varma}.
We have seen that this property arises from the electron
wavefunction renormalization, which contains in general nonlocal
contributions that can be studied rigorously within our
renormalization group approach. The logarithmic corrections to
the scale of the electron field are also consistent with another
known property, that is, the abnormal behavior of the
quasiparticle lifetime ($\sim \omega^{-1}$) in the model of the
van Hove singularity\cite{patt}. A standard Kramers-Kronig relation 
applied to the expression (\ref{self}) leads actually to an imaginary
part of the electron Green function linear in $\omega $, in
contrast to the conventional quadratic behavior in Fermi liquid
theory.

Finally, we have seen that 
the chemical potential itself also shows a nontrivial
scaling. At low energies, the Fermi level tends to be pinned
near the van Hove singularity. This flow of the chemical potential, 
unusual in condensed matter systems, may explain the existence
of a whole family of compounds where experiments show that the 
Fermi level is close to a van Hove singularity\cite{photo}.

We have not explored here the role played by the
intersingularity interactions, which may be important for the
mechanism driving to the superconducting phase. In fact, for a
sufficiently large coupling between the two singularities an
attractive channel may open, leading to pair binding with 
$d$-wave order parameter\cite{us}. It is worth noting that the Cooper
channel shows a $\log^2 \Lambda$ dependence on the cutoff, in
marked difference from the $\log \Lambda$ dependence usual in
the analysis of the Fermi liquid. In the context of our
renormalization group approach to the model, that instability
appears for a particular kinematics in the scattering process,
corresponding to the channel of bound state formation, and it is
therefore independent of the renormalization of the coupling
constant accomplished above. It would be interesting then to
undertake the analysis of the superconducting instability along
the lines followed in the present paper, to elucidate whether a
purely electronic mechanism may be responsible for the
superconductivity of the copper-oxide compounds.

\newpage

\section*{Appendix I}

The one-loop vertex function $i \Gamma^{(2)}$ in (\ref{vertex})
can be computed by letting the momentum integrals from $-\infty$
to $+\infty$, and carrying them out over the complex plane. In
the region with $\omega_q > 0$ and for $\omega > 0$, 
for instance, we  have
\begin{eqnarray}
I_{+}(\omega,k_x) & = & 
\int_{\omega_q > 0} d\omega_q \int d q_x d q_y
\frac{1}{\omega + \omega_q -
 t \left( (q_x + k_x)^2 - q^2_y \right) + 
  i\epsilon \: \rm{sgn}( \omega + \omega_q ) } \nonumber  \\
  &  &   \;\;\;\;\;\;\;\;\;\;\;\;\;\;\;\;\;\;\;\;\;\;  
  \times \frac{1}{\omega_q -
 t (q^2_x - q^2_y) + i\epsilon \: \rm{sgn} \: \omega_q }
                              \nonumber    \\
 &  = & - \frac{i \pi}{t}\int_{\omega_q > 0} d\omega_q \int d
  q_y \frac{1}{\omega - tk_x^2 - 2tk_x \sqrt{\omega_q /t + q_y^2
+ i \epsilon } + i \epsilon } \; \frac{1}{\sqrt{\omega_q /t + 
 q_y^2 + i \epsilon} }                 \nonumber      \\
 &   & - \frac{i \pi}{t}\int_{\omega_q > 0} d\omega_q \int d
  q_y \frac{1}{-\omega - tk_x^2 + 2tk_x \sqrt{(\omega + \omega_q) /t + q_y^2
+ i \epsilon } + i \epsilon }   \nonumber    \\
 &  &  \;\;\;\;\;\;\;\;\;\;\;\;\;\;\;\;\;\;\;\;\;\; \times 
   \frac{1}{\sqrt{ (\omega + \omega_q) /t + q_y^2 + i
 \epsilon } }               
\end{eqnarray}
after picking up the residue of the two poles on the upper
half-plane of the $q_x$ complex variable. The integrals over
$q_y$ can be done by deforming the contour of integration in
each case around a pole and a branch cut on the $q_y$ complex
plane. The sum of all the contributions gives
\begin{eqnarray}
I_{+}(\omega,k_x) & = & - \frac{i \pi}{t} f(\omega , k_x) 
\int_{1}^{\Lambda /(tk_x^2 f^2)} dx
 \frac{ {\rm arcsin} \sqrt{1 - 1/x} - \frac{\pi}{2} }
 { \sqrt{x - 1} }            \nonumber     \\
  &  &    + \frac{i \pi}{t} g(\omega , k_x) 
\int_{1}^{\Lambda /(tk_x^2 g^2)} dx
 \frac{ {\rm arcsin} \sqrt{1 - 1/x} - \frac{\pi}{2} }
 { \sqrt{x - 1} }            \nonumber     \\
  &  &    - \frac{i \pi}{t} g(\omega , k_x) 
\int_{0}^{\omega /(tk_x^2 g^2)} dx
 \frac{ {\rm arcsh} \sqrt{1/x - 1} }
 { \sqrt{1 - x} }            \nonumber     \\
  &  &  - \frac{\pi^2}{t} \left| f(\omega , k_x) \right|
 + \frac{\pi^2}{t} \left| g(\omega , k_x) \right| \label{mas}  \\
{\rm where} &  &  f(\omega,k_x) = \frac{\omega - tk_x^2}
  {2tk_x^2} \;\;\;\;\; {\rm and} \;\;\;\;\;
g(\omega,k_x) = \frac{\omega + tk_x^2}{2tk_x^2}  \nonumber 
\end{eqnarray}

The part $I_{-}$ of the one-loop integral for $\omega_q < 0$ can
be obtained in similar fashion, with the result that
\begin{eqnarray}
I_{-}(\omega,k_x) & = &  \frac{i \pi}{t} f(\omega , k_x) 
\int^{0}_{-\Lambda /(tk_x^2 f^2)} dx
 \frac{ {\rm arcsh} \sqrt{- 1/x} }
 { \sqrt{1 - x} }            \nonumber     \\
  &  &    - \frac{i \pi}{t} g(\omega , k_x) 
\int^{0}_{-\Lambda /(tk_x^2 g^2)} dx
 \frac{ {\rm arcsh} \sqrt{- 1/x} }
 { \sqrt{1 - x} }            \nonumber     \\
  &  &    + \frac{i \pi}{t} g(\omega , k_x) 
\int_{0}^{\omega /(tk_x^2 g^2)} dx
 \frac{ {\rm arcsh} \sqrt{1/x - 1} }
 { \sqrt{1 - x} }            \nonumber     \\
  &  &  - \frac{\pi^2}{t} \left| f(\omega , k_x) \right|
        +  \frac{\pi^2}{t} \left| g(\omega , k_x) \right|  
\label{menos}
\end{eqnarray}

Both expressions (\ref{mas}) and (\ref{menos}) combine to
produce a well-defined extension to the whole  $(k_x, k_y)$ 
plane realizing the SO(1,1) invariance of the model, as quoted
in the main text.

\newpage

\section*{Appendix II}

In order to compute the terms of ${\rm Re} \: \Sigma^{(2)} ({\bf
 p}, \omega_p)$ linear in $\varepsilon ({\bf p})$, we divide the
momentum integral in $(k_x,k_y)$ space in the four different
sectors of positive and negative energy separated by the Fermi
line $\varepsilon ({\bf k}) = 0$. In the region $k_y \geq \left|
k_x \right|$, for instance, we differenciate with respect to
$p_x^2$ and set the external momentum to zero (whenever it is
not needed to regulate an infrared divergence). Thus we obtain
\begin{eqnarray}
\lefteqn{\frac{\partial }{\partial p_x^2} {\rm Re} \:
\Sigma^{(2)}_{{\cal R}_1} ({\bf p} \rightarrow {\bf 0},0) =} 
                         \nonumber     \\
 & = & - \frac{1}{16 \pi^4} \frac{U^2}{t} \int^{\Lambda} dr \:
\frac{1}{r} \int_{0}^{\infty} dx \sum_{n = \pm 1} 
 \frac{x - n(1-\epsilon^2) + 2n}{\sqrt{\left( (x - n(1-\epsilon^2))^2
+ 4\epsilon^2 \right)^3 } }    \nonumber       \\
  &  & \;\;\;\;\;\;\;\;\;\;\;\;\;\;\;\;\;\;\;\; \times 
  \left( -\log \: \left|x^2 - 1\right|
+ x \: \log \: \left| \frac{x-1}{x+1} \right| \right) \nonumber \\
  &  & - \frac{1}{4 \pi^3}  \int^{\Lambda} dr \:
\frac{1}{r} \int_{0}^{\infty} dx \sum_{n = \pm 1} 
 \frac{x - n(1-\epsilon^2) + 2n}{\sqrt{ \left( (x - n(1-\epsilon^2))^2
+ 4\epsilon^2 \right)^3 } }          \nonumber     \\
  &  &  \;\;\;\;\;\;\;\;\;\;\;\;\;\;\;\;\;\;\;\; \times  
  \log \left( \frac{2 \epsilon}{x -
 n(1-\epsilon^2) +  \sqrt{ (x - n(1-\epsilon^2))^2
+ 4\epsilon^2  } } \right)  {\rm Im} \: \left. \Gamma^{(2)} 
 \right|_{\omega /\left|\varepsilon ({\bf k})\right| = x} \nonumber \\
  &  & + \frac{1}{8 \pi^3}  \int^{\Lambda} dr \:
\frac{1}{r} \int_{0}^{\infty} dx \sum_{n = \pm 1}
 \frac{n}{\epsilon^2} 
 \frac{1}{\sqrt{  (x - n(1-\epsilon^2))^2 + 4\epsilon^2  } }   
 \; {\rm Im} \: \left. \Gamma^{(2)} 
 \right|_{\omega /\left|\varepsilon ({\bf k})\right| = x} \nonumber \\
  &  & - \frac{1}{4 \pi^3}  \int^{\Lambda} dr \:
\frac{1}{r} \int_{0}^{\infty} dx \sum_{n = \pm 1} 
 \frac{x - n(1-\epsilon^2) + 2n}{ (x - n(1-\epsilon^2))^2
+ 4\epsilon^2  }          \nonumber     \\
  &  &   \;\;\;\;\;\;\;\;\;\;\;\;\;\;\;\;\;\;\;\; \times      
\frac{1}{x -  n(1-\epsilon^2) +  \sqrt{ (x - n(1-\epsilon^2))^2
+ 4\epsilon^2  } } \;  {\rm Im} \: \left. \Gamma^{(2)} 
 \right|_{\omega /\left|\varepsilon ({\bf k})\right| = x} \nonumber \\
  &  & - \frac{1}{4 \pi^3}  \int^{\Lambda} dr \:
\frac{1}{r} \int_{0}^{\infty} dx \sum_{n = \pm 1} 
 \frac{1}{\sqrt{  (x - n(1-\epsilon^2))^2
+ 4\epsilon^2  } }          \nonumber     \\
  &  &   \;\;\;\;\;\;\;\;\;\;\;\;\;\;\;\;\;\;\;\; \times      
\frac{1}{x -  n(1-\epsilon^2) +  \sqrt{ (x - n(1-\epsilon^2))^2
+ 4\epsilon^2  } } \;  {\rm Im} \: \left. \Gamma^{(2)} 
 \right|_{\omega /\left|\varepsilon ({\bf k})\right| = x}
\end{eqnarray}
where $\epsilon \equiv p_x/r$.
It can be checked that, when the contributions from the four
different sectors in $(k_x,k_y)$ space are put together, all the
$1/\epsilon^2$ terms cancel out. The leading contribution in
$\log \epsilon$ from the integrals in the $x$ variable may be evaluated
analytically. The contribution independent of $\epsilon$ is more
difficult to extract, and it has been obtained by numerical
computation of the above integrals, yielding the result quoted
in the main text.

\newpage
\section*{Figure Captions}
\mbox{  }

\noindent
{\bf Figure 1:} Contour map of the dispersion relation for a
two-dimensional square lattice with next-to-nearest-neighbor
hopping. 

\noindent
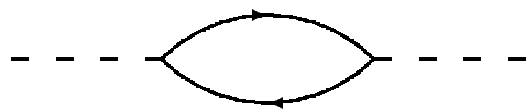
{\bf Figure 2:} One-loop order correction to the interaction
potential. 

\noindent
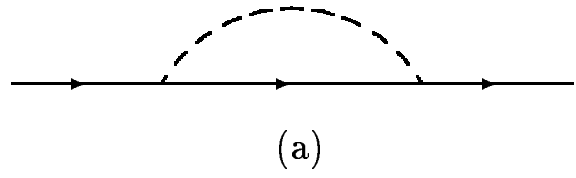
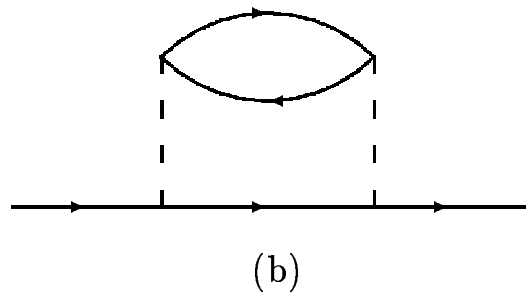
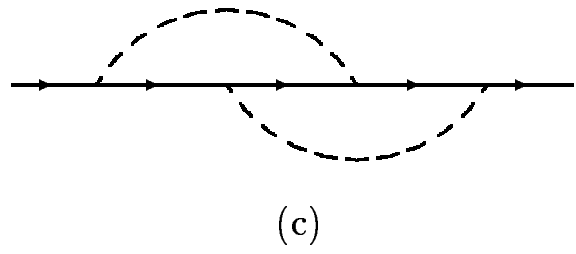
{\bf Figure 3:} Different self-energy diagrams to the two-loop
order. 

\noindent
{\bf Figure 4:} Plot of the beta function at the right-hand-side
of (\ref{flow}).

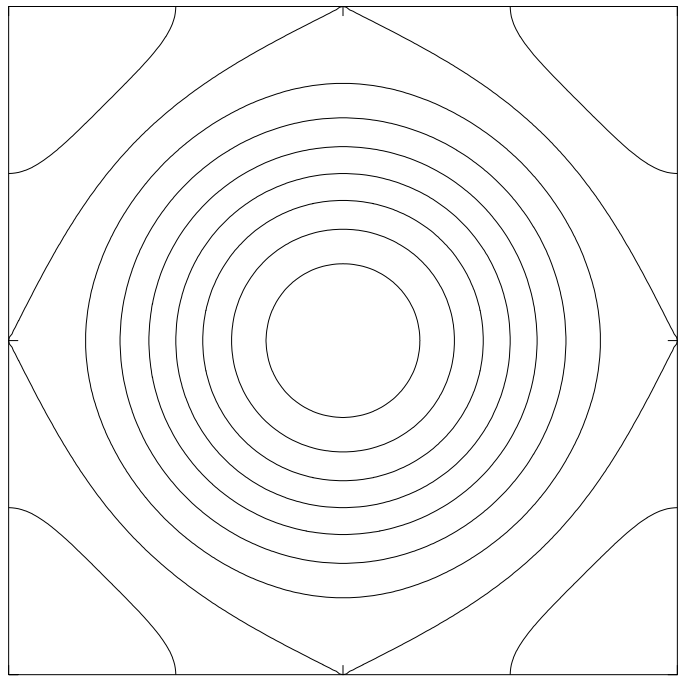
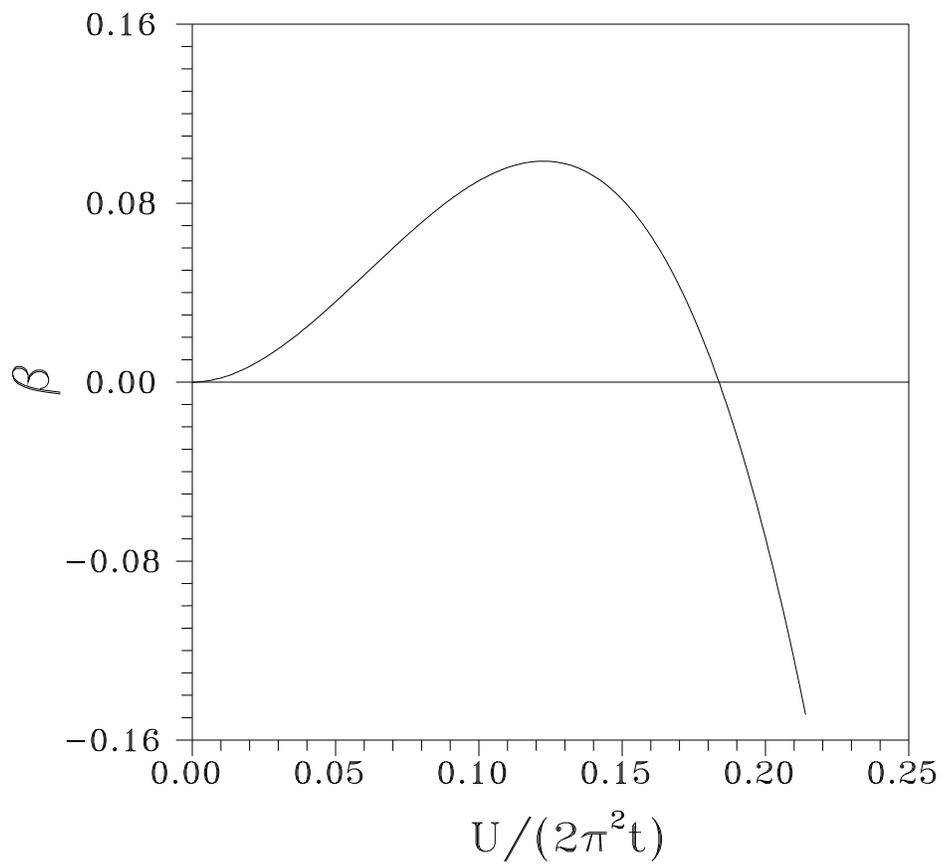
\end{document}